\documentclass[12pt,reqno]{amsart}

\usepackage{amssymb,amsthm,amsmath,mathrsfs,}
\usepackage{enumerate}
\usepackage{fullpage}
\usepackage{graphicx}
\usepackage{float}
\usepackage{subfig}

\begin{document}

\newtheorem{theorem}{Theorem}[section]
\newtheorem{lemma}[theorem]{Lemma}
\newtheorem{define}[theorem]{Definition}
\newtheorem{remark}[theorem]{Remark}
\newtheorem{corollary}[theorem]{Corollary}
\newtheorem{example}[theorem]{Example}
\newtheorem{assumption}[theorem]{Assumption}
\newtheorem{proposition}[theorem]{Proposition}
\newtheorem{conjecture}[theorem]{Conjecture}

\def\Ref#1{Ref.~\cite{#1}}

\def\i{\mathrm{i}}
\def\Rnum{{\mathbb R}}
\def\const{\text{const.}}
\def\sgn{\mathrm{sgn}}

\def\smallbinom#1#2{{\textstyle \binom{#1}{#2}}}
\def\smallsum{\textstyle\sum}
\def\smalliint{{\textstyle\iint}}

\def\pr{{\rm pr}}
\def\X{\mathbf{X}}
\def\Y{\mathbf{Y}}

\def\d{{\mathbf d}}
\def\id{\mathrm{id}}
\def\lieder#1{{\mathcal L}_{#1}}
\def\Esp#1{{\mathcal{E}_{#1}}}

\def\grad{{\boldsymbol\nabla}}
\def\lapl{{\boldsymbol\Delta}}

\def\U{Q}
\def\V{A}
\def\H{\sigma}
\def\B{\gamma}
\def\K{\kappa}

\def\ext{\text{ext.}}
\def\mov{\text{mov.}}
\def\s{\text{s}}
\def\c{\text{c}}

\tolerance=50000
\allowdisplaybreaks[3]

\title{Exact solutions and conservation laws\\ of a one-dimensional PDE model\\ for a blood vessel}

\author{
Stephen C. Anco$^1$,
Tamara M. Garrido$^2$,\\
Almudena P. M\'arquez$^2$,
Mar\'ia L. Gandarias$^2$
\\\\
${}^1$D\lowercase{\scshape{epartment}} \lowercase{\scshape{of}} M\lowercase{\scshape{athematics and}} S\lowercase{\scshape{tatistics}}\\
B\lowercase{\scshape{rock}} U\lowercase{\scshape{niversity}}\\
S\lowercase{\scshape{t.}} C\lowercase{\scshape{atharines}}, ON L2S3A1, C\lowercase{\scshape{anada}} \\
\\
${}^2$D\lowercase{\scshape{epartment}} \lowercase{\scshape{of}} M\lowercase{\scshape{athematics}}\\
U\lowercase{\scshape{niversity of}} C\lowercase{\scshape{adiz}}\\
11510 P\lowercase{\scshape{uerto}} R\lowercase{\scshape{eal}}, C\lowercase{\scshape{adiz}}, S\lowercase{\scshape{pain}}\\
}


\begin{abstract}
Two aspects of a widely used 1D model of blood flow in a single blood vessel are studied
by symmetry analysis, 
where the variables in the model are the blood pressure
and the cross-section area of the blood vessel. 
As one main result, 
all travelling wave solutions are found by
explicit quadrature of the model. 
The features, behaviour, and boundary conditions
for these solutions are discussed. 
Solutions of interest include shock waves and sharp wave-front pulses
for the pressure and the blood flow.
Another main result is that three new conservation laws
are derived for inviscid flows.
Compared to the well-known conservation laws in 1D compressible fluid flow,
they describe generalized momentum
and generalized axial and volumetric energies.
For viscous flows, these conservation laws get replaced by
conservation balance equations which contain a dissipative term
proportional to the friction coefficient in the model. 
\end{abstract}

\maketitle

\section{Introduction}

In recent years, 
one-dimensional (1D) models of blood flow in human blood vessels 
have been widely used in clinical applications \cite{MarWilLac2009,AudBucBekVibVig-CleGer2017}. 
These models are effective for understanding averaged features of blood flow locally, 
such as velocity, volume flux, and pressure \cite{SherForPeiFran2003,ForLamQua2003}.
They can also be combined with 3D models 
for detailed simulation of the human cardiovascular system as a whole 
\cite{ForMouNob2007,XiaAlaFig2013,DobLiaPanVas2019}. 
Moreover, 1D models have much less computational cost compared to 3D models
and can be mathematically analyzed in greater depth. 

A non-branching blood vessel in a 1D model is a cylindrical tube whose radius varies 
as a function of time $t$ and axial distance $x$,
in which the blood is an incompressible fluid governed by the Navier-Stokes equations 
averaged over cross-sections of the tube.
The variables consist of the cross-section area $A$, 
the volume flux $Q$ of blood flow, the mean pressure $P$, 
and the mean blood velocity $\bar u=Q/A$,
while the blood density is taken to be constant. 
$A$ and $Q$ satisfy a system of two coupled partial differential equations (PDEs)
which are similar in form to the Navier-Stokes equations 
for mass continuity and momentum in fluid mechanics.
The system is closed by specifying a relation for 
the pressure in terms of the cross-section area; 
the simplest widely-used model is that the pressure change across the vessel wall is proportional to the change in radius of the vessel. 
There are two important parameters in the resulting closed system: 
a friction parameter, 
which is proportional to the viscosity coefficient in the Navier-Stokes equations; 
and a momentum correction parameter, 
which arises from how the Navier-Stokes are averaged over a cross-section
\cite{BarHunTimVar}. 

In the literature, there is a lot of work on numerical solutions, 
but very little has been done on exact solutions 
except for steady-states \cite{MuMenSosFa,GhiBerLeTor2020,BriXin2020} 
and the use of the well-known Riemann method of characteristics 
for wave propagation \cite{SpiTorVazCon2017,SheZhaZhe2020}. 
The latter method, however, can be carried out to obtain explicit solutions 
only in a special case for the momentum correction parameter
\cite{ForLamQua2003}. 

The main purpose of the present paper is to illustrate the utility of symmetry analysis 
applied to a 1D model of blood flow in a single blood vessel, 
which will provide explicit analytical information about 
exact non-steady solutions and conservation laws. 
Two main results will be obtained. 

Firstly, all explicit travelling wave solutions of the model will be derived -- namely, 
a wave form for $A$ and $Q$ that moves with a constant speed and preserves its shape.
A complete discussion of these solutions and their properties is given.

One type of solution obtained describes a dissipating shock wave
in a very long constricting blood vessel with a steady-state near each end;
the vessel's diameter, pressure, and blood flow display a rapid transition
in the shock, which moves at a constant speed.
A similar shock wave solution is found for a blood vessel of arbitrary length
in which the initial state of the blood vessel is close to a steady-state
and then rapidly transitions such that
the diameter, pressure, and blood flow are increasing. 
The blood velocity exhibits a shock behaviour between two steady-states.

Another type of solution obtained describes a pulse with a sharp front
for the blood vessel's diameter, pressure, and blood flow; 
behind the front, these quantities decrease to a steady-state behaviour.
Other solutions are obtained that exhibit a similar sharp front,
with different behaviours behind the front.

In general, 
solutions that describe travelling waves in an infinitely long tube
can be applied to modelling a very long blood vessel
where the morphology at the ends of the vessel is not relevant. 
When the morphology at the ends is important,
an explanation will be given of how the travelling wave solutions
are applicable with various boundary conditions satisfied at the end points.
More discussion of the applicability of travelling waves will be given
at the end of the paper. 

Secondly, some explicit new conservation laws admitted by the model will be derived. 
The only well-known conservation laws to-date have been
the total blood volume and the net blood flux, 
plus an Eulerian energy quantity which holds only in a special case of 
the momentum correction parameter \cite{ForLamQua2003}. 
Three new conservation laws are obtained: 
a generalized momentum and two generalized energies,
which hold for any value of the momentum correction parameter 
and for a general pressure-area relation 
when the blood flow is modelled as being inviscid.
The momentum quantity is a modified form of the well-known momentum in fluid mechanics. 
The energy quantities represent a volumetric energy and an axial energy,
which are similar to generalizations of the well-known energy in fluid mechanics
(for the Euler equations of inviscid flow).
It is interesting, however, that there are two different conserved energies
for the blood flow model.
When the blood flow is modelled as being viscous,
then these three new quantities are no longer conserved
but they satisfy conservation balance equations that contain
a dissipative volume term proportional to friction coefficient.
Such balance equations are useful in mathematical analysis of the initial-value problem.

In particular, it is known that the Eulerian energy quantity 
satisfies a balance equation leading to a time-decay inequality \cite{ForLamQua2003}.
This quantity coincides with the total volumetric energy
in the blood flow model in a special case for the momentum correction parameter,
but otherwise it is not conserved in the blood flow model even for inviscid flow.
Analogous inequalities can be derived for the viscous blood flow model 
by use of the volumetric energy and axial energy
with no restriction on the momentum correction parameter.

Furthermore,
the new conservation laws yield associated boundary conditions for the model
such that the flux of the generalized momentum and the generalized energies is zero
in the rest frame of the blood flow. 
These zero-flux boundary conditions can be applied to travelling wave solutions,
as well as steady-state solutions. 

All of these results are new.
A worthwhile remark is that the viewpoint here is applied mathematics,
rather than biological modelling.

Section~\ref{sec:model} 
summarizes the blood flow PDEs 
and various commonly used forms for the pressure-area relation. 
The kinematical symmetries and the five basic conservation laws of the model
with a general pressure-area relation are derived.

Section~\ref{sec:bc}
has a general discussion of boundary conditions for the model,
including zero-flux boundary conditions coming from the three new conservation equations.

Section~\ref{sec:travelling.waves}
starts with deriving the system of ordinary differential equations (ODEs) satisfied by
travelling wave solutions of the blood flow PDEs.
These ODEs have a quadrature which is obtained
for the separate cases of inviscid and viscous flow. 
Spatial domains and boundary conditions for the solutions are then considered.
Some main features of travelling waves are shown to hold for all pressure-area relations.

Section~\ref{sec:solns} 
presents all of the exact travelling wave solutions
in the case of the most widely used pressure-area relation, 
and discusses their basic mathematical and physical features. 

Section~\ref{sec:remarks}
makes some concluding remarks.

\section{Summary and features of the 1D model}\label{sec:model}

The PDE system for the quantities $A(x,t)$, $Q(x,t)$, $P(x,t)$ 
describing cross-section area, blood flow, and pressure in a cylindrical blood vessel 
is given by \cite{BarHunTimVar}
\begin{align}
& A_t + Q_x =0,
\label{eqn1}
\\
& Q_t + \alpha (Q^2/A)_x + \rho_0^{-1}A P_x + k Q/A =0
\label{eqn2}
\end{align}
where $\alpha\geq 1$ is a momentum correction coefficient 
(determined by the axial velocity profile),
$k\geq 0$ is the friction coefficient (proportional to the viscosity). 
Here $\rho_0>0$ is the blood density, which is constant. 

The pressure-area relation, 
which closes the system and is sometime called a ``tube law'', 
has the general form \cite{VasSalSim} 
\begin{equation}\label{eos.general}
P = \beta f(A/A_0) +P_\ext
\end{equation}
where $\beta>0$ is a constant,
and $P_\ext$ is the external pressure caused by the tissue surrounding the blood vessel,
which will be assumed to be constant. 
If there is no change in pressure across the vessel wall, 
then the blood vessel is assumed to have a constant area $A=A_0$,
whereby $P(A_0)=P_\ext$ implies that $f(1)=0$. 
Physiologically, it is expected that the pressure change 
should be an increasing positive function of the area:
\begin{equation}\label{f.props}
f(A/A_0)\geq 0
\quad\text{ and }\quad 
f'(A/A_0)>0
\end{equation}
for $A>0$. 
A variety of functional relations with these properties have been proposed in the literature, 
for example \cite{ForLamQua2003,VasSalSim,SaNi2019,GhiBerLeTor2020}:
\begin{subequations}\label{P.A.rels}
\begin{align}
f & =\sqrt{A/A_0} -1; 
\label{P.A.radius}\\
f & =1-1/\sqrt{A/A_0};\\
f & =\exp(A/A_0 -1)-1; \\
f & =(A/A_0)^m -(A/A_0)^n,
\quad
m>0\geq n .
\end{align}
\end{subequations}
The most commonly used relation is \eqref{P.A.radius} which models 
the pressure change being proportional to the change in radius of the blood vessel, 
\begin{equation}\label{eos}
P - P_\ext = \beta (\sqrt{A/A_0} -1). 
\end{equation}

For details of the derivation of this model \eqref{eqn1}, \eqref{eqn2}, \eqref{eos}
and further explanation of its biological and physical features, 
see \Ref{QuaFor2004,BriXin2020,Kri2022a}. 

Substitution of the general pressure-area relation \eqref{eos.general} 
into the PDEs \eqref{eqn1}--\eqref{eqn2} 
yields the closed system 
\begin{align}
& A_t + Q_x =0,
\label{Aeqn}
\\
& Q_t + \alpha (Q^2/A)_x + \beta_0 (A/A_0) f'(A/A_0) A_x + k Q/A =0
\label{Qeqn}
\end{align}
where $\beta_0=\beta/\rho_0 >0$. 
In terms of $Q$ and $A$, the mean blood flow velocity is 
\begin{equation}\label{u}
\bar u = Q/A.
\end{equation}
This system, with $A>0$, is well known to be hyperbolic, 
and consequently it possesses two Riemann invariants 
which propagate with speeds 
\begin{equation}
c_\pm = \alpha \bar u \pm \sqrt{\beta_0 (A/A_0) f'(A/A_0) + \alpha(\alpha-1)\bar u^2} 
\end{equation}
(see e.g.\ \Ref{QuaFor2004}). 
This means that the system admits nonlinear waves that travel along the paths 
determined by $dx/dt = c_\pm$. 

The most important parameter in the system \eqref{Aeqn}--\eqref{Qeqn} 
is the friction coefficient $k\geq0$. 
In applications, 
there are two main cases of interest. 

\textbf{Viscous}: $k>0$. 
In this case, the system \eqref{Aeqn}--\eqref{Qeqn} is dissipative. 
As an illustration, 
spatially homogeneous solutions $(A(t),Q(t))$ satisfy $A'=0$ and $Q'=-kQ/A$,
which gives $A=A_\s$ and $Q=Q_0e^{-\tfrac{k}{A_s} t}$,
where $A_\s$ is a positive constant and $Q_0$ is an arbitrary constant.
These solutions describe a blood vessel with a constant radius and a blood flux that 
exponentially drops to zero on a time scale $A_s/k$. 

\textbf{Inviscid}: $k=0$. 
In this case, the system \eqref{Aeqn}--\eqref{Qeqn} is non-dissipative. 
Spatially homogeneous solutions $(A(t),Q(t))$ simply are constants, 
$A=A_\s>0$ and $Q=Q_\s$, 
which describes a blood vessel with a constant radius carrying a constant blood flux.

\subsection{Kinematic point symmetries}

The system \eqref{Aeqn}--\eqref{Qeqn} has the following 
kinematic transformation groups of symmetries:
\begin{align}
\text{space reflection}\quad  &
x\to -x,\
Q\to -Q
\label{x-refl.symm}
\\
\text{time translation}\quad & 
t\to t+\epsilon 
\label{t-transl.symm}
\\
\text{axial translation}\quad & 
x\to x+\epsilon
\label{x-transl.symm}
\\
\text{scaling}\quad & 
t\to e^{\epsilon} t,\
x\to e^{(1+\frac{q}{2})\epsilon} x,\
A\to e^{\epsilon} A,\
Q\to e^{(1+\frac{q}{2})\epsilon} Q,
\quad
f = (A/A_0)^q -1
\label{scal.symm}
\end{align}
where $\epsilon$ is the parameter in the symmetry group. 
In the inviscid case, 
the system has additional kinematic symmetry transformation groups:
\begin{align}
\text{time reversal}\quad  &
t\to -t,\
Q\to -Q
\label{t-revers.symm}
\\
\text{dilation}\quad & 
t\to e^{\epsilon} t,\
x\to e^{\epsilon} x
\label{dil.symm}
\\
\text{Galilean boost}\quad & 
x\to x + \epsilon t,\
Q\to Q + \epsilon A,
\quad 
\alpha =1
\label{galilean.symm}
\end{align}
Note that the Galilean boost corresponds to $\bar u\to\bar u +\epsilon$. 

These symmetries \eqref{x-refl.symm}--\eqref{galilean.symm} are evident by 
inspection of the form of the system \eqref{Aeqn}--\eqref{Qeqn}
in comparison to the 1D Navier-Stokes equations for compressible fluids 
whose Lie point symmetries and discrete symmetries are well known
\cite{Ovs-book,Ibr-book,JiwKumSin}.

A determination of all point transformation symmetries is fairly complicated 
and will involve utilizing the form of the Riemann invariants. 

For a general discussion of symmetries and their applications to PDEs, 
see \Ref{Ovs-book,Olv-book,BCA-book}.

\subsection{Basic conservation laws}

Some conservation laws of the system \eqref{Aeqn}--\eqref{Qeqn}
can be readily found by comparison with
the well-known conservation laws of mass, momentum, and energy in the inviscid case, 
for 1D compressible fluid dynamics \cite{Ovs-book,Ibr-book,WebZan}. 
Mathematically, 
$A$ is analogous to mass density $\rho$, 
and $Q$ is analogous to the momentum density $\rho u$, 
where $\rho$ and $u$ are density and velocity variables 
in the 1D inviscid fluid equations
\begin{equation}
\rho_t + (\rho u)_x =0,
\quad
(\rho u)_t + (\rho u^2 + p(\rho))_x =0 ,
\end{equation}
while the fluid pressure $p(\rho)$ then corresponds to
$\int_{A_0}^A \beta_0 (A/A_0) f'(A/A_0) \, dA$. 
This analogy is exact in the case $k=0$, $\alpha =1$.

Firstly, 
the PDE \eqref{Aeqn} itself is a continuity equation for $A$ viewed as a density. 
Integration of $A$ over any portion $x_1\leq x\leq x_2$ of a blood vessel 
gives the total volume of blood in that portion:
\begin{equation}\label{vol}
V = \int_{x_1}^{x_2} A\,dx.
\end{equation}
This quantity satisfies the conservation law
\begin{equation}\label{vol.conslaw}
\tfrac{d}{dt} V = -Q\big|_{x_1}^{x_2}
\end{equation}
stating that the rate of change in blood volume is balanced by 
the net change in blood flow through the end points $x=x_1$ and $x=x_2$. 
The analogous conservation law in 1D fluid dynamics is the mass. 

Likewise, 
the PDE \eqref{Qeqn} in the inviscid case, $k=0$, is a continuity equation
for $Q$ viewed as a density. 
The integral of $\tfrac{1}{L}Q$ over $x_1\leq x\leq x_2$, with $L=x_2-x_1$, 
gives the net (mean) blood flux 
\begin{equation}\label{flux}
\bar Q = \tfrac{1}{L}\int_{x_1}^{x_2} Q\,dx 
\end{equation}
which satisfies the conservation law
\begin{equation}\label{flux.conslaw}
\tfrac{d}{dt} \bar Q = 
-\tfrac{1}{L}\big( \alpha Q^2/A + \beta_0 \big(A f(A/A_0) - F(A)\big) \big)\big|_{x_1}^{x_2}
= -\tfrac{2}{L}\rho_0^{-1} {\bar P} A \big|_{x_1}^{x_2}  
\end{equation}
with $F(A)= \int_{A_0}^A f(A/A_0)\,dA$, 
where 
\begin{equation}
\bar P = \tfrac{1}{2}\alpha\rho_0 {\bar u}^2 + \tfrac{1}{2} \big(P(A) - \beta A^{-1} F(A) -P_\ext\big)
\end{equation}
is the analog of mechanical pressure in inviscid constant-density fluid dynamics
\cite{Bat-book}.
Thus, the rate of change in the net blood flux is proportional to
the difference in the mechanical force ${\bar P}A$
on the cross-sections at each end.
In the case of viscous blood flow,
the conservation law is replaced by a balance equation
\begin{equation}
\tfrac{d}{dt} \bar Q = -\tfrac{2}{L}\rho_0^{-1} {\bar P} A \big|_{x_1}^{x_2}  - k\bar U 
\end{equation}
where $\bar U = \tfrac{1}{L}\int_{x_1}^{x_2} \bar u\, dx$ is mean velocity. 
This is analogous to the momentum balance equation in 1D viscous fluid dynamics. 

Secondly, through the analogy mentioned earlier, 
energy in 1D inviscid fluid dynamics corresponds to the quantity \cite{ForLamQua2003}
$E = \int_{x_1}^{x_2} (\tfrac{1}{2} Q^2/A + \beta_0 F(A))\,dx$, 
which satisfies 
$\tfrac{d}{dt} E = -\big( Q^3/A^2 + \beta_0 Q f(A/A_0) \big|_{x_1}^{x_2}$
when $\alpha=1$ and $k=0$. 
This energy balance equation can be derived directly 
from the summed product of the PDEs \eqref{Aeqn}--\eqref{Qeqn}
and the pair of expressions 
$\big({-\tfrac{1}{2}}Q^2/A^2 + \beta_0 A^2 f(A/A_0),Q/A\big)$, 
called a multiplier. 

A generalized energy can be sought for $\alpha\neq 1$ 
by adjusting the multiplier using a suitable power of $A$. 
Specifically, the adjusted multiplier 
$\big(({-\tfrac{r}{2}}Q^2 + h(A))A^{-1-r} ,Q A^{-r}\big)$ 
with a suitable choice of $h(A)$ leads to the balance equation
\begin{equation}\label{gen.ener.eqn}
\big( \tfrac{1}{2} Q^2 A^{-r}  +\beta_0 J(A;r) \big)_t 
+ \big( \tfrac{4\alpha-r}{6} Q^3 A^{-r-1} +\beta_0 Q J(A;r) \big)_x 
=- k Q^2 A^{-r-1}
\end{equation}
with
\begin{equation}\label{J}
J(A;r)= A H(A;r) +(r-2) \int_{A_0}^A H(A;r)\,dA,
\quad
H(A;r) = \int_{A_0}^A A^{-r} f(A/A_0)\,dA, 
\end{equation}
where $r^2 -(4\alpha-1)r + 2\alpha=0$. 
The roots 
\begin{equation}\label{r.vals}
r_\pm = 2\alpha -\tfrac{1}{2} \pm 2\sqrt{\alpha(\alpha-1) +\tfrac{1}{16}}
\end{equation}
are real and satisfy $r_+ \geq 2$, $1 \geq r_- >\tfrac{1}{2}$ since $\alpha \geq 1$. 
In the inviscid case, $k=0$, 
integration of the density term in the balance equation \eqref{gen.ener.eqn}
times $\rho_0$ gives the integral quantities
\begin{equation}\label{gen.ener}
E^\pm = \int_{x_1}^{x_2} \big( \tfrac{1}{2} \rho_0 Q^2/A^{r_\pm} +\beta J(A;r_\pm) \big) \,dx
\end{equation}
which satisfy the conservation laws
\begin{equation}\label{gen.ener.conslaw}
\tfrac{d}{dt} E^\pm
= - \big( \tfrac{4\alpha-r_\pm}{6}\rho_0 Q^3/A^{r_\pm+1}  + \beta Q H(A;r_\pm) \big)\big|_{x_1}^{x_2} . 
\end{equation}
In the viscous case, 
the righthand side of the conservation equation \eqref{gen.ener.conslaw}
will also contain a dissipative integral term 
$-k\int_{x_1}^{x_2} Q^2/A^{r_\pm+1} \,dx$. 

For $\alpha=1$, note that the generalized energy integrals \eqref{gen.ener} become 
\begin{equation}
E^- = \rho_0\int_{x_1}^{x_2}( \tfrac{1}{2}\bar u^2 + \beta_0 J^-(A) )A\,dx = E
\quad\text{ and }\quad
E^+ = \rho_0\int_{x_1}^{x_2}( \tfrac{1}{2} \bar u^2 +\beta_0 J^+(A) )\,dx
\end{equation}
in terms of 
$J^-(A) =J(A;r_-) = \int_0^1 f(\lambda A/A_0)\,d\lambda$
and 
$J^+(A) = J(A;r_+) = A\int_{A_0}^A A^{-2}f(A/A_0)\,dA$
after simplifications, 
where $r_-=1$, $r_+=2$. 
The quantity $E^-$ is the volumetric energy of the blood flow, 
while the other quantity $E^+$ is the axial energy. 

The same method also leads to a generalized momentum 
which arises from the multiplier $((1-2\alpha) Q A^{-2\alpha},A^{1-2\alpha})$.
This yields a balance equation 
\begin{equation}
\big( Q A^{1-2\alpha} \big)_t + \big( \tfrac{1}{2} Q^2 A^{-2\alpha} +\beta_0 G(A;\alpha) \big)_x 
=-k Q A^{-2\alpha} 
\end{equation}
with 
\begin{equation}\label{G}
G(A;\alpha) = A^{2(1-\alpha)} f(A/A_0) + 2(\alpha-1) \int_{A_0}^A A^{1-2\alpha} f(A/A_0)\,dA .
\end{equation}
Integration of the density term times $\rho_0$ gives the integral quantity  
\begin{equation}\label{gen.mom}
M = \rho_0  \int_{x_1}^{x_2} Q/A^{2\alpha-1}\,dx 
= \rho_0  \int_{x_1}^{x_2} \bar u/A^{2(\alpha-1)}\,dx
\end{equation}
satisfying 
\begin{equation}\label{gen.mom.conslaw}
\tfrac{d}{dt} M
= - \big( \tfrac{1}{2}\rho_0 Q^2/A^{2\alpha} + \beta G(A;\alpha) )\big|_{x_1}^{x_2}
-k\int_{x_1}^{x_2} Q/A^{2\alpha} \,dx . 
\end{equation}
This conservation equation becomes a conservation law in the inviscid case, $k=0$. 

For $\alpha=1$, note that 
the generalized momentum integral \eqref{gen.mom} 
reduces to $M=\rho_0 \int_{x_1}^{x_2} \bar u\,dx$
which is the momentum of the blood flow. 
It is interesting that, in contrast to 1D fluid dynamics, 
the blood flow system possesses this additional conserved momentum 
as well as an additional conserved energy $E^+$. 

A determination of all low-order conserved integrals and balance equations is, 
in principle, possible by the method of multipliers. 
However, similarly to the situation for symmetries, 
it is fairly complicated and will involve utilizing the form of the Riemann invariants. 

For a general discussion of multipliers and conservation laws of PDEs, 
see \Ref{Olv-book,BCA-book,AncBlu2002b,Anc-review}.

\subsection{Conserved integrals moving with the flow}

In fluid dynamics, 
it is useful to formulate conservation laws on domains that move with the fluid flow
\cite{Ibr-book,Bat-book}. 
A similar formulation can be given for 
the conservation equations \eqref{vol.conslaw}, \eqref{flux.conslaw}, 
\eqref{gen.mom.conslaw} and \eqref{gen.ener.conslaw}
so that they hold on moving domains $x_1(t)\leq x(t)\leq x_2(t)$ 
in the blood flow, as given by
\begin{equation}\label{moving}
\frac{d x(t)}{dt} = \bar u(t,x(t)) . 
\end{equation}

The moving blood volume is defined by 
\begin{equation}\label{vol.moving}
V_\mov = \int_{x_1(t)}^{x_2(t)} A\,dx 
\end{equation}
which satisfies 
\begin{equation}\label{vol.movingconslaw}
\tfrac{d}{dt} V_\mov = 0 . 
\end{equation}
Thus $V_\mov$ is a constant of motion. 
The moving net blood flux
\begin{equation}\label{flux.moving}
\bar Q_\mov = \frac{1}{L(t)}\int_{x_1(t)}^{x_2(t)} Q\,dx,
\end{equation}
with $L(t)=x_2(t)-x_1(t)$, 
satisfies the conservation law
\begin{equation}\label{flux.movingconslaw}
\tfrac{d}{dt}\big( L(t) \bar Q_\mov \big) = 
-\big( (\alpha-1) Q^2/A + \beta_0 \big(A f(A/A_0) - F(A)\big) \big)\big|_{x_1(t)}^{x_2(t)}
- kL(t)\bar U_\mov
\end{equation}
where $\bar U_\mov = \tfrac{1}{L(t)}\int_{x_1(t)}^{x_2(t)} \bar u\, dx$ is 
the moving mean velocity. 

In the inviscid case, 
the moving version of the conservation balance equations \eqref{gen.mom.conslaw} and \eqref{gen.ener.conslaw} 
take the form 
\begin{equation}\label{gen.mom.movingconslaw}
\tfrac{d}{dt} M_\mov
= - \big( \beta G(A;\alpha) - \tfrac{1}{2}\rho_0 Q^2/A^{2\alpha} )\big|_{x_1(t)}^{x_2(t)}
\end{equation}
for the moving generalized momentum 
\begin{equation}\label{gen.mom.moving}
M_\mov = \rho_0  \int_{x_1(t)}^{x_2(t)} \bar u/A^{2(\alpha-1)}\,dx, 
\end{equation}
and
\begin{equation}\label{gen.ener.movingconslaw}
\tfrac{d}{dt} E^\pm_\mov
= - \big( \beta Q G(A;\tfrac{1}{2}r_\pm)  -\tfrac{1}{6}(r_\pm +3 -4\alpha)\rho_0 Q^3/A^{r_\pm+1} \big)\big|_{x_1(t)}^{x_2(t)}
\end{equation}
for the moving generalized volumetric energy $(-)$ and axial energy $(+)$
\begin{equation}\label{gen.ener.moving}
E^\pm_\mov = \int_{x_1(t)}^{x_2(t)} \big( \tfrac{1}{2} \rho_0 Q^2/A^{r_\pm} +\beta J(A;r_\pm) \big) \,dx ,
\end{equation}
where $r_\pm$ is given by expression \eqref{r.vals} in terms of $\alpha$.

\section{Boundary conditions}\label{sec:bc}

As a model for blood flow, 
the system \eqref{Aeqn}--\eqref{Qeqn} must be supplemented 
by boundary conditions at the ends of blood vessel, 
$x=x_1$ and $x=x_2$, with $x_2>x_1$. 
Because this system is hyperbolic, 
a general argument based on the theory of characteristics indicates that 
a single boundary condition can be posed at each end 
\cite{DobLiaPanVas2019,GriKar2008,AlaParShe-review}. 
The specific type of boundary condition involves 
the particular biological morphology of the ends of the blood vessel being modelled: 
an end that is branching; 
an end that terminates or is blocked; 
an end that is open;
an end that has blood pumped in or out;
an end that has a fixed diameter or a fixed pressure; 
an end at which a pressure wave or blood flow pulse is propagating in or out; 
an end with a steady-state pressure.
Attention here will be restricted to the latter two cases.
Note that, depending on the morphology, 
the two ends can have different types of boundary conditions. 

Propagation of a pressure wave with speed $c$ along a blood vessel
is specified by conditions $cP_x = P_t$ at each end.
From the pressure-area relation \eqref{eos.general},
this is equivalent to the boundary conditions
\begin{equation}\label{bc:A.wave}
cA_x(x_1,t) =A_t(x_1,t),
\quad 
cA_x(x_2,t) =A_t(x_2,t),
\quad
t\geq 0.
\end{equation}
Likewise, boundary conditions specifying a blood flow pulse are given by
\begin{equation}\label{bc:Q.wave}
cQ_x(x_1,t) =Q_t(x_1,t),
\quad 
cQ_x(x_2,t) =Q_t(x_2,t),
\quad
t\geq 0.
\end{equation}

For modelling a very long blood vessel, the ends can be regarded as being at 
$x_1\to -\infty$ and $x_2\to \infty$. 
Boundary conditions are thereby regarded as holding asymptotically. 
A precise meaning of very long is that the total length $x_2 -x_1$
is much greater than any length scale
in the equations \eqref{Aeqn} and \eqref{Qeqn} 
and in the initial conditions $A(x,0)$, $Q(x,0)$.

\subsection{Zero-flux boundary conditions}

Another kind of boundary condition can be obtained from 
the flux terms in the conservation equations \eqref{gen.mom.movingconslaw} and \eqref{gen.ener.movingconslaw}
for the generalized momentum and the generalized energies 
on a domain $x_1(t)\leq x(t)\leq x_2(t)$ moving with the blood flow
(cf \eqref{moving}). 

Setting the generalized momentum flux expression to vanish
yields the following conditions: 
\begin{equation}\label{mom.zeroflux}
\tfrac{1}{2} Q(x_1,t)^2 = \beta_0 A(x_1,t)^{2\alpha} G(A(x_1,t);\alpha),
\quad
\tfrac{1}{2} Q(x_2,t)^2 = \beta_0 A(x_2,t)^{2\alpha} G(A(x_2,t);\alpha),
\quad
t\geq 0
\end{equation}
where $G(A;\alpha)$ is non-negative when $\alpha\geq 1$ and $A\geq A_0$, 
as seen from by expression \eqref{G} since $f$ is a non-negative function. 
The meaning of this boundary condition \eqref{mom.zeroflux} is that
the moving generalized momentum \eqref{gen.mom} of the blood flow 
is conserved for a solution $(A(x,t),Q(x,t))$ in the inviscid case.
In the viscous case, the meaning is that the moving generalized momentum
for a solution exhibits dissipation with no flux.

Likewise, setting the flux expression of the generalized energies 
to vanish yields the following conditions: 
\begin{equation}\label{ener.zeroflux}
\begin{aligned}
& \tfrac{1}{6}(r_\pm +3 -4\alpha) Q(x_1,t)^2 
 = \beta_0 A(x_1,t)^{r_\pm+1} G(A(x_1,t);\tfrac{1}{2}r_\pm) , 
\quad
t\geq 0
\\
& \tfrac{1}{6}(r_\pm +3 -4\alpha) Q(x_2,t)^2 
 = \beta_0 A(x_2,t)^{r_\pm+1} G(A(x_2,t);\tfrac{1}{2}r_\pm)  ,
\quad
t\geq 0 .
\end{aligned}
\end{equation}
It is straightforward to show from expression \eqref{r.vals} that the coefficient 
$r_\pm +3 -4\alpha$ is positive for $\alpha \geq 1$ in the $+$ case; 
in the $-$ case, 
this coefficient is $0$ for $\alpha =1$ and decreases for large values of $\alpha$. 
Therefore, 
since $G(A;\tfrac{1}{2}r_\pm)$ is non-negative when $\alpha\geq 1$ and $A\geq A_0$, 
the ``$+$'' boundary condition is consistent. 
It has the meaning that the moving generalized axial energy of the blood flow, 
given by the integral \eqref{gen.ener.moving} in the $+$ case, 
is conserved for a solution $(A(x,t),Q(x,t))$ in the case of inviscid flow, 
while it exhibits dissipation with no flux for a solution
in the case of viscous flow. 
The ``$-$'' boundary condition, 
which would have a similar meaning in terms of the moving generalized volumetric energy, 
is consistent only for $A\leq A_0$.

\section{Travelling waves}\label{sec:travelling.waves}

A travelling wave has the form 
\begin{equation}\label{travelwave}
A =\V(\xi),
\quad
Q=\U(\xi),
\quad
\xi=x-ct
\end{equation}
where $c$ is the wave speed.
This form arises from group-invariance with respect to the translation symmetry
$(t,x)\to (t+\epsilon,x+c\epsilon)$,
with group parameter $\epsilon$. 

If $c=0$, then a travelling wave reduces to a steady-state solution. 
Hereafter, $c$ will be taken to be non-zero.
Substitution of expressions \eqref{travelwave} 
into the blood flow system \eqref{Aeqn}--\eqref{Qeqn} 
yields the travelling wave ODEs
\begin{equation}
-c \V' + \U' =0,
\quad
-c \U' + \alpha (\U^2/\V)' + \beta_0 (\V/\V_0) f'(\V/\V_0) \V' + k \U/\V =0.
\end{equation}
The first ODE gives $\U$ in terms of $\V$,
and then the second ODE becomes a nonlinear separable equation for $\V$:
\begin{equation}
\U=c\V + C_1,
\quad
\big( c^2(\alpha-1) + \beta_0 (\V/\V_0) f'(\V/\V_0) - C_1^2 \alpha \V^{-2} \big)\V' +ck + C_1 k\V^{-1} =0.
\end{equation}
Let 
\begin{equation}\label{rels}
C=-C_1/c,
\quad
\gamma = (\alpha -1) c^2 \geq 0, 
\quad 
\sigma = C^2\alpha c^2 >0,
\quad
\kappa=k c \neq 0.
\end{equation}
The equations for $\U$ and $\V$ now have the simpler form  
\begin{equation}\label{U.eqn}
\U=c(\V-C)
\end{equation}
and 
\begin{equation}\label{V.eqn.gen}
\V' = \frac{\kappa(\V -C)\V}{\sigma - \gamma \V^2 - (\beta_0/A_0) \V^3 f'(\V/\V_0)}.
\end{equation}
Note that the physical parameters are given in terms of $\kappa$, $\sigma$, $\gamma$ by the relations
\begin{equation}\label{params}
\alpha = \H/(\H -\B C^2),
\quad
c = \pm\sqrt{\H -\B C^2}/(\sqrt{2} C),
\quad
k = \pm \K C/(\sqrt{2}\sqrt{\H -\B C^2}).
\end{equation}
Also note that the properties \eqref{f.props} of a general pressure-area function
$f(A/A_0)$ imply that $\V^3 f'(\V/\V_0) \geq 0$ for $\V\geq 0$. 

Some general features of solutions in the inviscid and viscous cases
will be discussed next.

\subsection{Inviscid flow}
When $k=0$, equation \eqref{V.eqn.gen} for $\V(\xi)$ reduces to $\V'=0$.
Hence, $\V$ is constant, 
and consequently equation \eqref{U.eqn} shows that $\U$ is also constant. 
These two constants determine the value of $C=A-Q/c$. 

Thus, the general solution is a homogeneous steady state:
\begin{equation}
\V =\V_s =\const >0,
\quad
\U=\U_\s=\const
\end{equation}
The mean blood flow velocity is $\bar u = \bar u_s= \U_\s/\V_\s$,
while the pressure is $P=P_\s = P_\ext + \beta f(A_\s/A_0)$. 
In these steady states, there is no pulsatility of the blood flow.
Physiologically, this describes an equilibrium state, 
which is called a “living-human equilibrium” in the literature 
\cite{BriXin2020}.

\subsection{Viscous flow}
For $k>0$, 
equation \eqref{V.eqn.gen} gives a quadrature for $\V(\xi)$. 
Up to a shift in $\xi$, there is a one-parameter family of solutions $\V(\xi)$
in terms of the arbitrary constant $C$. 
The features of the solution family depend essentially on the sign of $C$
and on the value of the positive root of the denominator in the quadrature. 
Let \begin{equation}\label{A1}
\V_\c>0,
\quad
\gamma \V_\c^2 +(\beta_0/A_0) \V_\c^3 f'(\V_\c/\V_0) -\sigma =0
\end{equation}
If $\V_c\neq C$, then 
an asymptotic expansion of equation \eqref{V.eqn.gen} for $\V$ near $\V_\c$
shows that $\xi$ is finite and thus it is the location of a one-sided cusp 
where
\begin{equation}
\V(\xi_\c) = \V_\c,
\quad
\V'(\xi_\c)=\infty.
\end{equation} 
For $\V$ near $0$, an asymptotic expansion of equation \eqref{V.eqn.gen} 
shows that $|\xi|\to\infty$, which thus represents an exponential tail in $\V$. 
Similarly, if $\V_\c\neq C>0$, then $\V$ near $C$ has an exponential tail. 

Attention will be restricted to solutions with positive wave speeds, $c>0$. 
Solutions with negative wave speed are given by reflection $\xi\to-\xi$ 
applied to positive-wave speed solutions,
since $\kappa$ changes sign while $\sigma$ and $\gamma$ are invariant
under $c\to -c$.

\subsection{Domain and boundary conditions}

Firstly,
consider a travelling wave solution $\V(\xi)$ on $-\infty<\xi<\infty$.
This corresponds to a solution
\begin{equation}\label{AQ.travelwave}
(A,Q)=(\V(x-ct),c(\V(x-ct)-C))
\end{equation}  
of the system \eqref{Aeqn}--\eqref{Qeqn}
on the spatial domain $-\infty<x<\infty$,
where the asymptotic behaviour of $\V(\xi)$
determines the type of asymptotic boundary conditions
holding for the solution $(A,Q)$. 

If $\V(\xi)$ asymptotically approaches a steady-state,
then $(A,Q)$ will satisfy asymptotic steady-state boundary conditions
\begin{equation}\label{bc:A.asympt.ss}
A_x\to 0,
\quad
Q_x\to 0,
\quad\text{as}\quad
x\to \pm\infty,
\quad
t\geq 0.
\end{equation}
If $\V(\xi)$ has other asymptotic behaviour, then $(A,Q)$ will satisfy
asymptotic wave propagation boundary conditions
\begin{equation}\label{bc:A.asympt.wave}
cA_x - A_t \to 0,
\quad
cQ_x - Q_t \to 0 
\quad\text{as}\quad
x\to \pm\infty,
\quad
t\geq 0
\end{equation}
since travelling waves \eqref{AQ.travelwave} automatically satisfy such boundary conditions at any point $x$.

Secondly,
consider a travelling wave solution $\V(\xi)$
on only a finite domain $\xi_1\leq \xi\leq \xi_2$.
This will yield a corresponding solution \eqref{AQ.travelwave}
of the system \eqref{Aeqn}--\eqref{Qeqn}
on a finite spatial domain $x_1\leq x\leq x_2$
in a finite time interval $0\leq t\leq T$ 
which are given as follows.

Suppose $c>0$.
At $t=0$, the front of the wave will define
the location of the right end point $x=x_2$ via the relation $\xi_2=x_2$.
The left end point $x=x_1$ will be defined by
the location of the back of the wave at $t=T$ via $\xi_1=x_1-cT$.
The size of the domain is thus $x_2 - x_1= \xi_2-\xi_1 - cT$
which requires that $T<(\xi_2-\xi_1)/c$.
Thus, the point $\xi=\xi_2$ on the wave starts at $x=x_2$
and moves to the right, out of the spatial domain,
while the point $\xi=\xi_1$ on the wave starts out of the spatial domain
and moves to the right, entering the domain at $t=T$.
A similar discussion applies when $c<0$. 

At the end points of the domain $x_1\leq x\leq x_2$,
the solution \eqref{AQ.travelwave} will satisfy
wave propagation boundary conditions \eqref{bc:A.wave} or \eqref{bc:Q.wave}.

Thirdly,
consider a travelling wave solution $\V(\xi)$
on a half-infinite domain $-\infty < \xi\leq \xi_2$.
The corresponding solution \eqref{AQ.travelwave} 
of the system \eqref{Aeqn}--\eqref{Qeqn}
is defined for $t\geq 0$ on a spatial domain that can be 
either finite, $x_1\leq x\leq x_2$,
or half-infinite, $-\infty < x\leq x_2$. 

Finally, note that a travelling wave solution $\V(\xi)$
on the domain $-\infty<\xi<\infty$ 
can be truncated to any interval $\xi_1\leq \xi\leq \xi_2$
to obtain a solution \eqref{AQ.travelwave} on a finite domain. 

Apart from wave propagation boundary conditions,
it is possible to consider zero-flux boundary conditions
posed on the moving domain with respect to $\xi$.
This will be pursued elsewhere.

\section{Exact solutions}\label{sec:solns}

Based on the discussion in the previous section, 
travelling wave solutions \eqref{AQ.travelwave} 
for different pressure-area relations \eqref{P.A.rels} will be qualitatively similar. 
Here the simplest and most common pressure-area relation \eqref{eos}
will be considered, 
with the travelling wave ODE \eqref{V.eqn} having the explicit form
\begin{equation}\label{V.eqn}
\V' = \frac{\kappa(\V -C)\V}{\sigma - \gamma \V^2 - \mu \V^{5/2}} 
\end{equation}
where
\begin{equation}
\mu = \beta_0/(2\sqrt{A_0}) >0 . 
\end{equation}
All solutions $A(\xi)$ will now be presented,
and their detailed features and physical interpretation will be discussed.

\subsection{Solutions for $C=0$}

\begin{figure}[h]
\centering
\includegraphics[width=.40\textwidth]{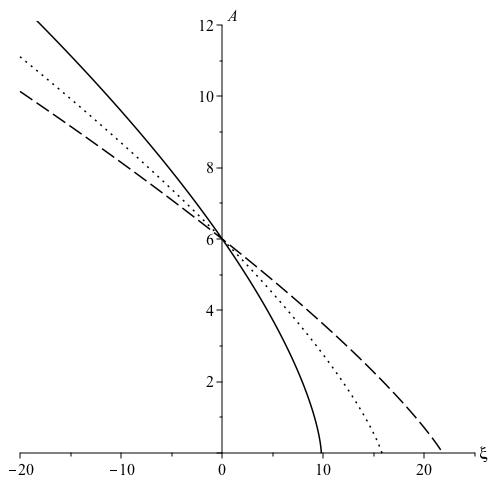}
\caption{$C=0$: area profile for 
$\gamma/\mu=$ $0$ (solid), $1$ (dot), $2$ (dash).}
\label{fig:Cis0-A}
\end{figure}

In this case, $\sigma=0$ from relations \eqref{rels}. 
The quadrature of equation \eqref{V.eqn} is then given by 
$\tfrac{2}{3} \mu \V^{3/2} +\gamma \V = \kappa(\xi_0-\xi)$,
where $\xi_0$ is an integration constant. 
This is a cubic equation for $\sqrt{\V}$ which can be solved explicitly. 
Solutions have the behaviour that $\V(\xi)$ is a concave decreasing function of $\xi$ 
that reaches zero at $\xi=\xi_0$ where $\V'(\xi_0)=-\kappa/\gamma<0$ 
from equation \eqref{V.eqn}. 
Equation \eqref{U.eqn} yields $\U(\xi)=c\V(\xi)$, 
and thus $\bar u = c$ is constant.
See Fig.~\ref{fig:Cis0-A}. 

Extending $\V(\xi)$ to be a piecewise solution that is $0$ past $\xi=\xi_0$,
then this describes a blood vessel 
that is filling behind the front $x=\xi_0+ct$ of a moving blood flow pulse 
and that is constricted ahead of the front,
with the blood flow velocity being the same as the speed of the front, $\bar u=c$.

\subsection{Solutions for $C<0$}

In this case, equation \eqref{V.eqn} has the quadrature
\begin{equation}\label{quadrature.Cisneg}
\begin{aligned}
& 
2\mu |C|^{3/2} \arctan\big(\sqrt{\V}/\sqrt{|C|}\big)
-(\sigma/|C|) \ln(\V) -(\gamma |C| - \sigma/|C|) \ln(\V+|C|) 
\\& 
+ \tfrac{2}{3} \mu \V^{3/2} +\gamma \V -2\mu |C| \sqrt{\V} = \kappa(\xi_0-\xi)
\end{aligned}
\end{equation}
which determines $\V(\xi)$.
By translation symmetry, it is convenient to put $\xi_0=0$,
which corresponds to a shift in either the $x$ or $t$ coordinates. 
Solutions exhibit the following two different behaviours,
which are distinguished by whether $\V(\xi) \gtrless \V_\c$. 

\begin{figure}[h]
\centering
\includegraphics[width=.40\textwidth]{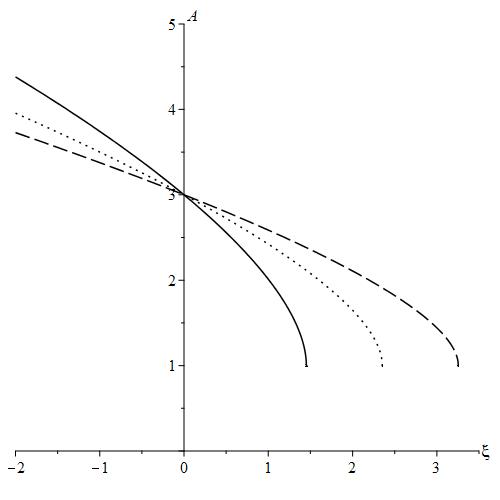}
\quad
\includegraphics[width=.40\textwidth]{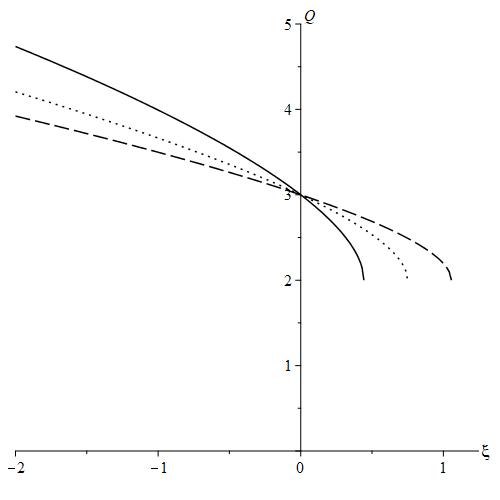}
\\
\includegraphics[width=.40\textwidth]{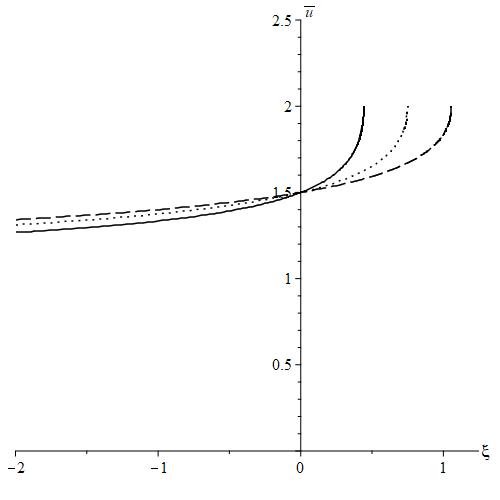}
\caption{$C<0$, $\V(\xi)>\V_\c$: area, blood flow, blood velocity profiles for 
$\gamma/\mu=$ $0$ (solid), $1$ (dot), $2$ (dash).}
\label{fig:Cisneg-VgtrV1}
\end{figure}

If $\V(\xi) >\V_\c$, then the solution $\V(\xi)$ is a concave decreasing function of $\xi$ 
that exhibits an inverted one-sided cusp at $\xi=\xi_\c$, 
and the solution does not exist for $\xi>\xi_\c$. 
From equation \eqref{U.eqn}, 
$\U(\xi)=c(\V(\xi)+|C|)$ has a similar behaviour, except for a constant offset. 
Thus, $\bar u(\xi) = c(1 + |C|/\V(\xi))$ is a positive, convex increasing function of $\xi$,
with a one-sided cusp at $\xi=\xi_\c$. 
See Fig.~\ref{fig:Cisneg-VgtrV1}. 

By extending $(\V(\xi),\U(\xi),\bar u(\xi))$ 
as a piecewise (continuous) solution that is constant past $\xi=\xi_\c$,
this describes a blood vessel 
that is expanding behind the sharp front $x=\xi_\c+ct$ of a moving blood flow pulse. 
The blood velocity exceeds the speed $c$ of the pulse ahead of the front, 
and dips down to the pulse speed $c$ far behind the front,
such that the rate of change spikes at the front, $\bar u_x|_{x=\xi_\c+ct}=\infty$. 

\begin{figure}[h]
\centering
\includegraphics[width=.40\textwidth]{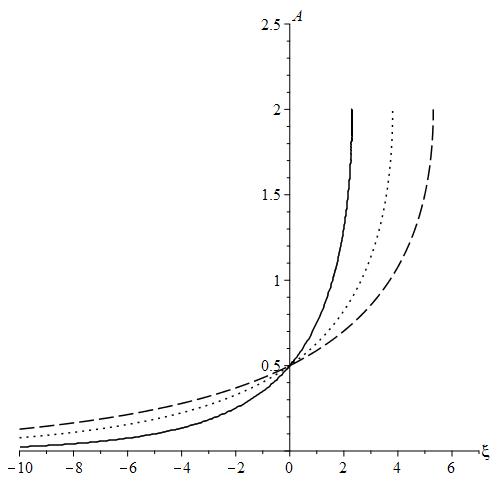}
\quad
\includegraphics[width=.40\textwidth]{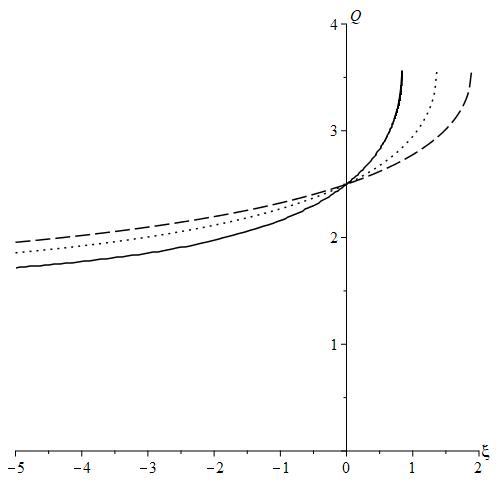}
\\
\includegraphics[width=.40\textwidth]{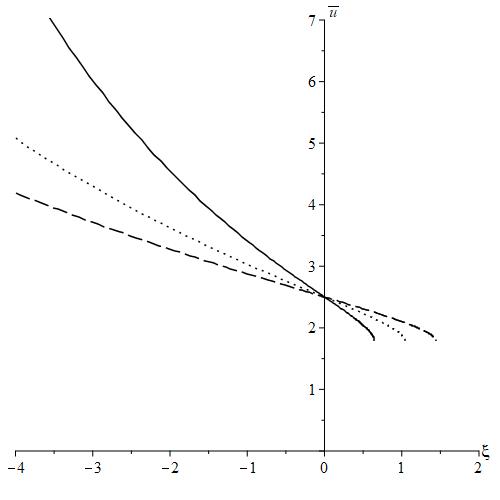}
\caption{$C<0$, $\V(\xi)<\V_\c$: area, blood flow, blood velocity profiles for 
$\gamma/\mu=$ $0$ (solid), $1$ (dot), $2$ (dash).}
\label{fig:Cisneg-V1gtrV}
\end{figure}

If $\V(\xi) <\V_\c$, then the solution behaviour is that 
$\V(\xi)$ goes to zero exponentially as $\xi\to -\infty$ 
and is a convex increasing function of $\xi$ with a one-sided cusp at $\xi=\xi_\c$. 
$\U(\xi)=c(\V(\xi)+|C|)$ again has a similar behaviour with a constant offset,
and $\bar u(\xi) = c(1 + |C|/\V(\xi))$ is a decreasing positive function of $\xi$ 
with an inverted at $\xi=\xi_\c$ where $\bar u(\xi_\c)=c(1+|C|/\V_\c)>0$. 
See Fig.~\ref{fig:Cisneg-V1gtrV}.

Extension of $(\V(\xi),\U(\xi),\bar u(\xi))$ 
as a piecewise (continuous) solution that is constant past $\xi=\xi_\c$ 
describes a blood vessel 
that is constricting behind the sharp front $x=\xi_\c+ct$ of a moving blood flow pulse. 
The blood velocity $\bar u$ exceeds the speed $c$ of the pulse ahead of the front, 
and rises behind the front, such that the rate of change spikes at the front,
$\bar u_x|_{x=\xi_\c+ct}=-\infty$.

\subsection{Solutions for $C>0$}

In this case,
the quadrature of equation \eqref{V.eqn} for $\V(\xi)$ is given by 
\begin{equation}\label{quadrature.Cispos}
\begin{aligned}
& 
\mu C^{3/2} \ln\big(|\sqrt{C} - \sqrt{\V}|/(\sqrt{C} +\sqrt{\V})\big)
+(\sigma/C) \ln(\V) +(\gamma C - \sigma/C) \ln(|C - \V|) 
\\& 
+ \tfrac{2}{3} \mu \V^{3/2} +\gamma \V + 2\mu C \sqrt{\V} = \kappa(\xi_0-\xi).
\end{aligned}
\end{equation}
Again, it is convenient to put $\xi_0=0$ by translation symmetry. 
Solutions exhibit several different behaviours,
which are distinguished by whether $\V(\xi) \gtrless \V_\c$ and  $\V(\xi) \gtrless C$
and also whether $\V_\c\gtrless C$, as follows. 

The solutions in the case $\V_\c>C$ will be discussed first. 

\begin{figure}[h]
\centering
\includegraphics[width=.40\textwidth]{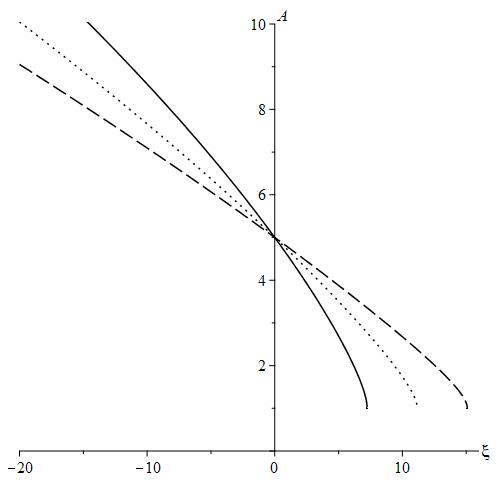}
\quad
\includegraphics[width=.40\textwidth]{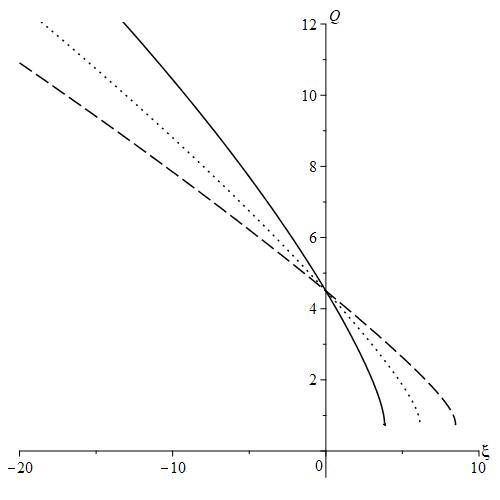}
\\
\includegraphics[width=.40\textwidth]{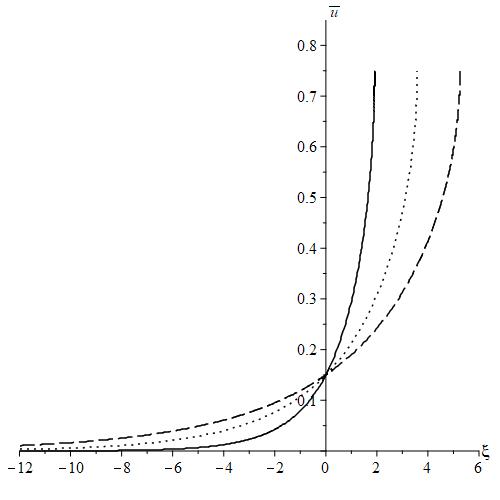}
\caption{$\V(\xi)>\V_\c>C>0$: area, blood flow, blood velocity profiles for 
$\gamma/\mu=$ $0$ (solid), $1$ (dot), $2$ (dash).}
\label{fig:Cispos-VgtrV1gtrC}
\end{figure}

Suppose $\V(\xi)>\V_\c>C$. 
The qualitative solution behaviour is similar to the corresponding case when $C<0$: 
$\V(\xi)$ and $\U(\xi)=c(\V(\xi)-C)$ are positive concave decreasing functions of $\xi$,
which have an inverted one-sided cusp at $\xi=\xi_\c$ where the solution stops. 
Thus, $\bar u(\xi)= c(1-C/\V(\xi))$ is a positive convex increasing function 
that exponentially approaches the value $0$ as $\xi\to -\infty$ and has a one-sided cusp at $\xi=\xi_\c$. 
See Fig.~\ref{fig:Cispos-VgtrV1gtrC}.

By extending $(\V(\xi),\U(\xi),\bar u(\xi))$ 
as a piecewise (continuous) solution that is constant past $\xi=\xi_\c$,
this describes a blood vessel 
that is expanding behind the sharp front $x=\xi_\c+ct$ of a moving blood flow pulse. 
The blood velocity $\bar u$ is less than the speed $c$ of the pulse ahead of the front, 
and drops to zero behind the front, such that there is a spike in the rate of change at the front, 
$\bar u_x|_{x=\xi_\c+ct}=\infty$. 

\begin{figure}[h]
\centering
\includegraphics[width=.40\textwidth]{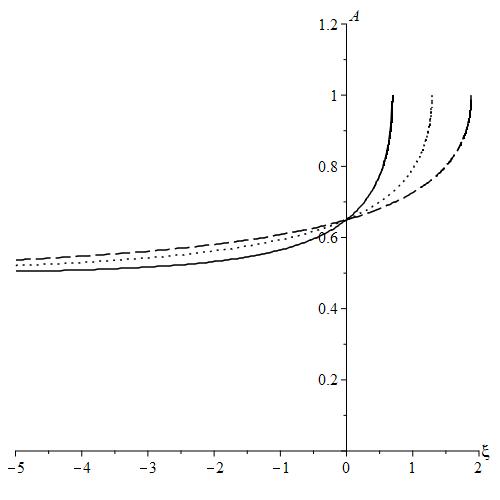}
\quad
\includegraphics[width=.40\textwidth]{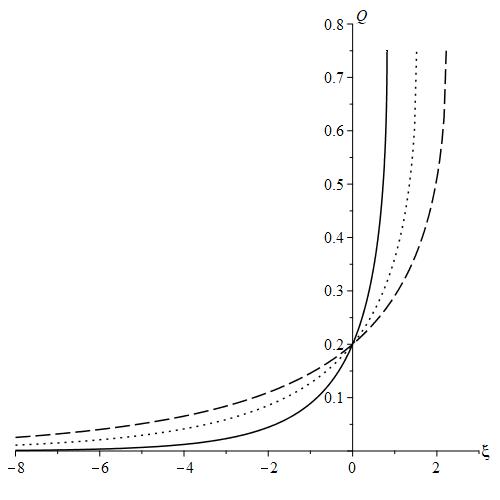}
\\
\includegraphics[width=.40\textwidth]{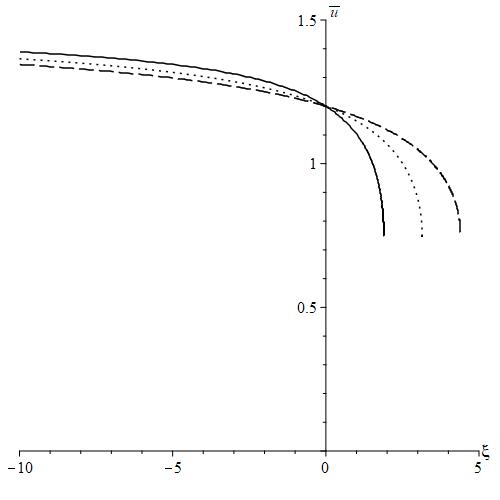}
\caption{$\V_\c>\V(\xi)>C>0$: area, blood flow, blood velocity profiles for 
$\gamma/\mu=$ $0$ (solid), $1$ (dot), $2$ (dash).}
\label{fig:Cispos-V1gtrVgtrC}
\end{figure}

Suppose $\V_\c>\V(\xi)>C$. 
The qualitative solution behaviour is again similar to the corresponding case when $C<0$: 
$\V(\xi)$, $\U(\xi)=c(\V(\xi)-C)$, and $\bar u(\xi)= c(1-C/\V(\xi))$ 
are positive increasing functions which exhibit a one-sided cusp at $\xi=\xi_\c$, 
and the solution does not exist for $\xi>\xi_\c$. 
As $\xi\to -\infty$, 
$\V(\xi)$ approaches the value $C>0$ exponentially,
while $\U(\xi)$ and $\bar u(\xi)$ go to zero. 
See Fig.~\ref{fig:Cispos-V1gtrVgtrC}.

Extension of $(\V(\xi),\U(\xi),\bar u(\xi))$ 
as a piecewise (continuous) solution that is constant past $\xi=\xi_\c$
describes a blood vessel that contracts sharply inward 
to a constant diameter behind the front $x=\xi_\c+ct$ of a moving blood pulse 
at which the rate of decrease in area and blood flow have a spike. 
The blood velocity $\bar u$ is less than the speed $c$ of the pulse ahead of the front, 
and rises slowly to the speed of the pulse behind the front, 
such that there is a spike in the rate of change at the front,
$\bar u_x|_{x=\xi_\c+ct}=-\infty$. 

\begin{figure}[h]
\centering
\includegraphics[width=.40\textwidth]{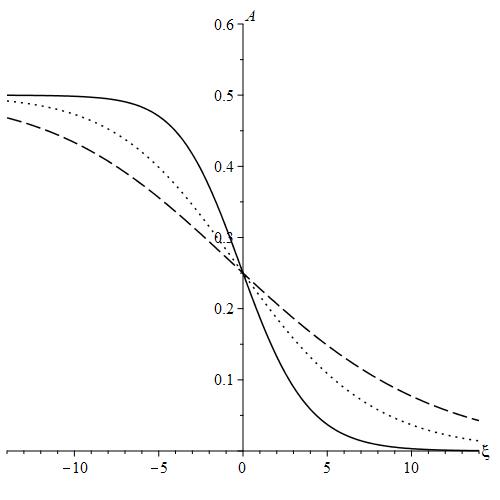}
\quad
\includegraphics[width=.40\textwidth]{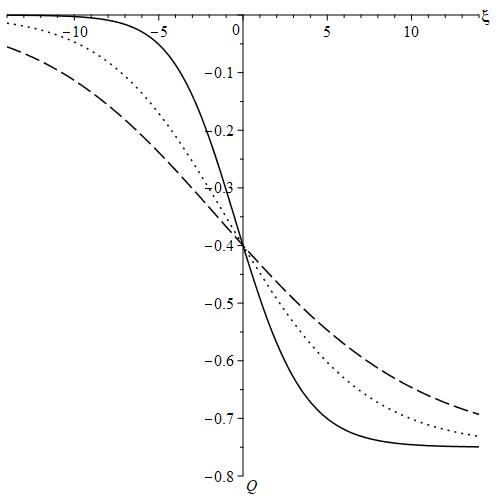}
\\
\includegraphics[width=.40\textwidth]{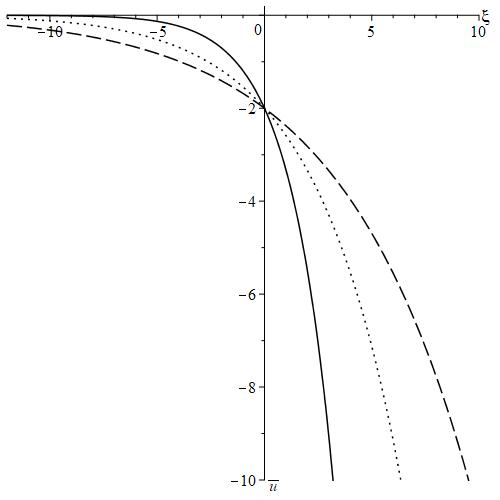}
\caption{$\V_\c>C>\V(\xi)>0$: area, blood flow, blood velocity profiles for 
$\gamma/\mu=$ $0$ (solid), $1$ (dot), $2$ (dash).}
\label{fig:Cispos-V1gtrCgtrV}
\end{figure}

For $\V_\c>C>\V(\xi)$, a different behaviour arises. 
The solution exists for all $\xi$ and has the asymptotic behaviour that 
$\V(\xi)$ decreases exponentially to $0$ as $\xi\to \infty$,
and exponentially approaches the value $C$ as $\xi\to -\infty$. 
Consequently, $\U(\xi)=c(\V(\xi)-C)$ is a negative function that 
exponentially approaches the value $0$ as $\xi\to -\infty$ 
and decreases to the value $-c C$ as $\xi\to \infty$. 
Therefore, $\bar u(\xi) = c(1 -C/\V(\xi))$ is a negative decreasing function 
that exponentially approaches the value $0$ as $\xi\to -\infty$ 
but has no lower bound as $\xi\to \infty$. 
See Fig.~\ref{fig:Cispos-V1gtrCgtrV}.

This describes a blood vessel in which there is a moving compressive pulse 
with a shock front that causes the blood flow to be in the backward direction. 
Far ahead of the pulse, the blood vessel is constricted such that the diameter is close to zero,
while at the front of the pulse, where convexity of the cross-section area vanishes, 
the blood flow exhibits a sharp transition from a high flow value to a low flow value. 

Next, the solutions in the case $C>\V_\c$ will be discussed. 

\begin{figure}[h]
\centering
\includegraphics[width=.40\textwidth]{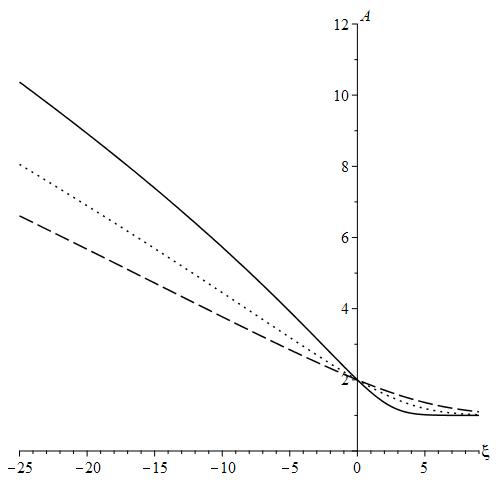}
\quad
\includegraphics[width=.40\textwidth]{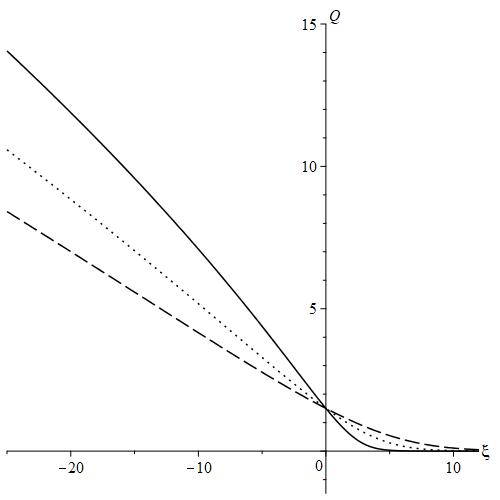}
\\
\includegraphics[width=.40\textwidth]{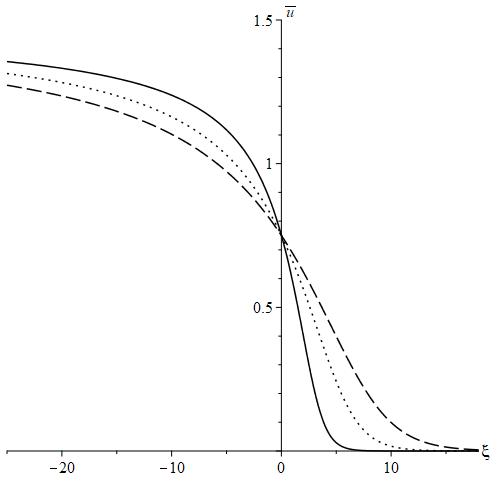}
\caption{$\V(\xi)>C>\V_\c>0$: area, blood flow, blood velocity profiles for 
$\gamma/\mu=$ $0$ (solid), $1$ (dot), $2$ (dash).}
\label{fig:Cispos-VgtrCgtrV1}
\end{figure}

Suppose $\V(\xi)>C$. 
The solution $\V(\xi)$ is a positive concave decreasing function of $\xi$ 
that exponentially approaches the value $C$ as $\xi\to \infty$,
while $\U(\xi)=c(\V(\xi)-C)$ has a similar behaviour but goes to $0$ as $\xi\to \infty$. 
Thus, $\bar u(\xi) = c(1 -C/\V(\xi))$ is a positive decreasing function 
that exponentially goes to $0$ as $\xi\to \infty$. 
See Fig.~\ref{fig:Cispos-VgtrCgtrV1}.

This describes a blood vessel whose diameter increases as the blood flows forward 
along the vessel. 
Unlike previous cases, the pulse does not have a sharp front,
but there is a transition point where the rate of decrease in blood flow and velocity 
reaches a maximum, 
with the velocity tapering to zero ahead of this point. 
The blood velocity resembles a shock whose front, where the convexity vanishes, 
corresponds to the transition point. 

\begin{figure}[h]
\centering
\includegraphics[width=.40\textwidth]{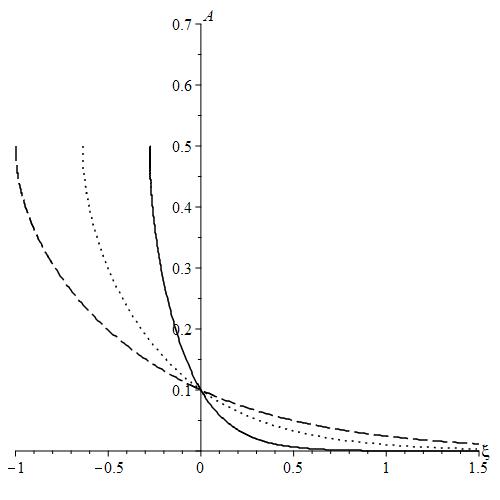}
\quad
\includegraphics[width=.40\textwidth]{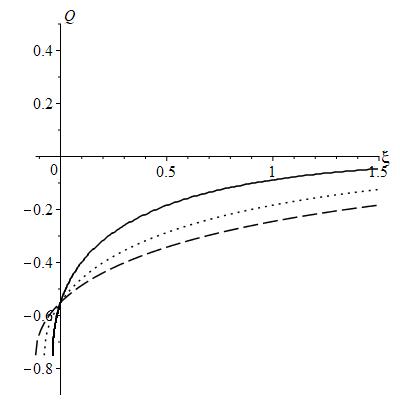}
\\
\includegraphics[width=.40\textwidth]{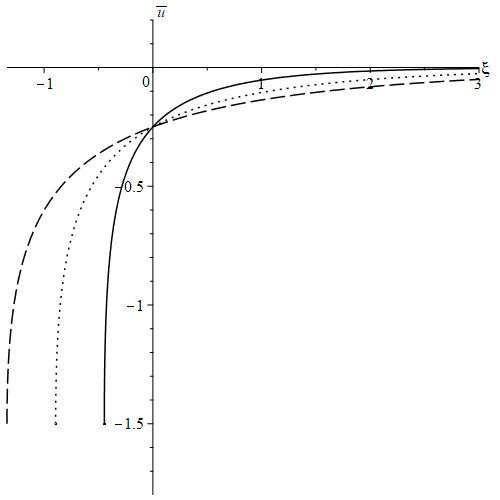}
\caption{$C>\V(\xi)>\V_\c>0$: area, blood flow, blood velocity profiles for 
$\gamma/\mu=$ $0$ (solid), $1$ (dot), $2$ (dash).}
\label{fig:Cispos-CgtrVgtrV1}
\end{figure}

Suppose $C>\V(\xi)>\V_\c$. 
The solution $\V(\xi)$ is a positive concave increasing function of $\xi$ 
that starts as an inverted one-sided cusp at $\xi=\xi_\c$ 
and exponentially approaches the value $C$ as $\xi\to \infty$. 
Likewise, $\U(\xi)=c(\V(\xi)-C)$ and $\bar u(\xi) = c(1 -C/\V(\xi))$ 
start with an inverted negative one-sided cusp at $\xi=\xi_\c$ and are increasing functions of $\xi$ 
that exponentially approach $0$ as $\xi\to \infty$. 
See Fig.~\ref{fig:Cispos-CgtrVgtrV1}.

Extending $(\V(\xi),\U(\xi),\bar u(\xi))$ 
as a piecewise (continuous) solution that is constant prior to $\xi=\xi_\c$,
then this describes a blood vessel in which there is a backward pulse of blood flow 
with a sharp front $x=\xi_\c+ct$ that moves forward at speed $c$. 
Ahead of the front, the vessel is constricted and the blood velocity $\bar u$ is close to zero, 
while at the front the diameter flares to a constant size $2\sqrt{A_1/\pi}$ 
and the backward velocity rapidly rises up to the speed $c$ of the pulse at the front. 

\begin{figure}[h]
\centering
\includegraphics[width=.40\textwidth]{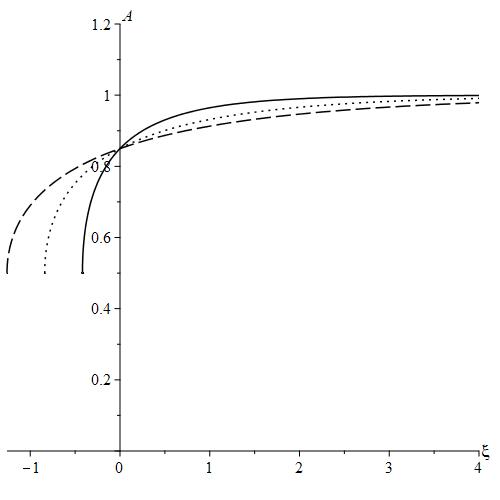}
\quad
\includegraphics[width=.40\textwidth]{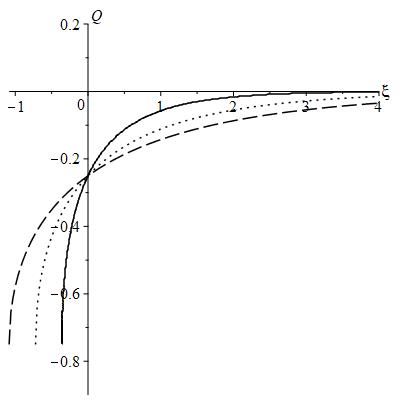}
\\
\includegraphics[width=.40\textwidth]{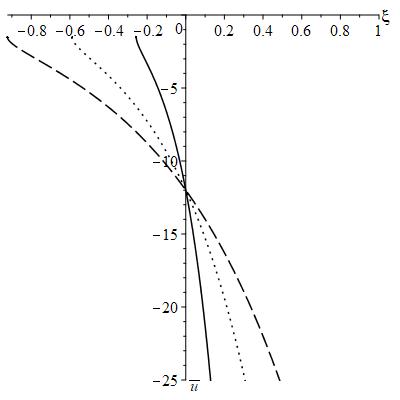}
\caption{$C>\V_\c>\V(\xi)>0$: area, blood flow, blood velocity profiles for 
$\gamma/\mu=$ $0$ (solid), $1$ (dot), $2$ (dash).}
\label{fig:Cispos-CgtrV1gtrV}
\end{figure}

A different behaviour occurs for $\V_\c>\V(\xi)$. 
The solution $\V(\xi)$ is a positive convex decreasing function of $\xi$ 
that starts as a one-sided cusp at $\xi=\xi_\c$ and exponentially approaches $0$ as $\xi\to \infty$. 
$\U(\xi)=c(\V(\xi)-C)$ starts as a negative one-sided cusp at $\xi=\xi_\c$
and decreases with $\xi$ such that it exponentially goes to $0$ as $\xi\to \infty$. 
$\bar u(\xi) = c(1 -C/\V(\xi))$ similarly is a negative decreasing function of $\xi$ 
that starts with a one-sided cusp at $\xi=\xi_\c$, but it has no lower bound as $\xi\to \infty$. 
See Fig.~\ref{fig:Cispos-CgtrV1gtrV}.

Extension of $(\V(\xi),\U(\xi),\bar u(\xi))$ 
as a piecewise (continuous) solution that is constant prior to $\xi=\xi_\c$
describes a blood vessel that rapidly collapses to a constant diameter $2\sqrt{A_1/\pi}$
behind the sharp front $x=\xi_\c+ct$ of moving pulse with speed $c$. 
The blood flow and velocity are in the backward direction
and their rate of change spikes at the front,
$Q_x|_{x=\xi_\c+ct}=-\infty$ and $\bar u_x|_{x=\xi_\c+ct}=\infty$. 
Ahead of the front, the blood flow tapers to zero 
while the blood velocity is rising in magnitude. 

Last, consider the case $\V_\c=C>0$.
The root equation \eqref{A1} yields $\V_c=\V_*$ where
\begin{equation}
\V_*=(\tfrac{2\rho_0 c^2}{\mu})^2,
\end{equation}
and equation \eqref{V.eqn} for $\V(\xi)$ becomes 
\begin{equation}
\V' = \frac{\kappa \V}{(\V+C)(\gamma +\mu \V^{3/2})}.
\end{equation}
Its quadrature is given by 
\begin{equation}\label{quadrature.CisV1}
\gamma \V_* \ln(\V) 
+ \tfrac{2}{3} \mu \V^{3/2} +\gamma \V + 2\mu \V_* \sqrt{\V} = \kappa(\xi_0-\xi).
\end{equation}

The solution behaviour in this case is similar to the earlier case $\V(\xi)>\V_\c>C$,
except that here $\V(\xi)$ goes to $0$ exponentially, 
$\U(\xi)=c(\V(\xi)-\V_*)$ changes sign and goes to $-c\V_*$ exponentially,
while $\bar u(\xi) = c(1 -\V_*/\V(\xi))$ changes sign and decreases with no lower bound. 

A useful final remark is that 
\begin{equation}\label{V1.C.cond}
\sgn(\V_\c -C)=\sgn(\V_* -C)
\end{equation}
can be shown to hold from the root equation \eqref{A1}
by the following argument. 
First, use the relations \eqref{rels} to write equation \eqref{A1} as
$\tilde\gamma \V_\c^2 +\tilde\mu \V_\c^{5/2}= C^2$
where $\tilde\gamma = 1-\tfrac{1}{\alpha}\geq0$ 
and $\tilde\mu = \tfrac{\mu}{2\alpha c^2}>0$
which do not involve $C$. 
Then, for $\V_\c=C\neq0$, 
$C=(\tfrac{1-\tilde\gamma}{\tilde\mu})^2 = \V_*$. 
Next, view $\V_\c-C:= f(C)$ as a function of $C$. 
The root equation shows that the solutions of $f(C)=0$ are $C=\V_*$ and $C=0$,
and that $f'(\V_*)= -\tfrac{1-\tilde\gamma}{5-\tilde\gamma} = -\tfrac{1}{4\alpha+1}$ 
is negative. 
This implies $f(C)>0$ for $0<C<\V_*$ and $f(C)<0$ for $C>\V_*$,
which establishes the sign relation \eqref{V1.C.cond}.

\section{Concluding remarks}\label{sec:remarks}

In the present work, 
symmetry analysis has been applied to a widely used 1D model of 
blood flow in a single blood vessel, with a general pressure-area relation. 
Several new results have been obtained. 

One main result is that 
three new conservation laws have been derived in case of inviscid flow.
These conservation laws yield conserved integrals describing
generalized momentum and generalized volumetric and axial energies.
The generalized momentum differs compared to the momentum
in inviscid constant-density 1D fluid dynamics by involving powers of $A$ that 
depend on $\alpha-1$, where $\alpha$ is the momentum correction coefficient. 
Likewise, when $\alpha\neq 1$,
both of the generalized energies involve different powers of $A$
compared to the energy in inviscid constant-density fluid dynamics.
In the case of viscous blood flow,
each conservation law gets replaced by a balance equation 
containing a dissipative volume term proportional to the friction coefficient
in the model.

These conservation laws can be expected to be useful in analysis of the model
\cite{ForLamQua2003}.
In particular, they can provide conserved norms and enable the derivation of 
time-decay inequalities; 
they can also be used for checking the accuracy of numerical schemes. 

Another main result is that travelling wave solutions have been studied in detail. 
Prior to this contribution, the only exact solutions which have been studied 
in the literature were steady-state solutions. 
Firstly, general features of the travelling wave solutions have been discussed 
and shown to be qualitatively independent of the specific form of 
the pressure-area relation in the model. 
Secondly, all travelling waves have been derived explicitly for the 
simplest and most commonly considered case where 
the pressure change across the blood vessel wall is proportional to the change in radius. 
These solutions are most naturally applicable to the idealized case of
a very long blood vessel in which the morphology of ends is not relevant,
as the spatial domain of a solution in this situation is unbounded.
For a blood vessel whose morphology at the ends is important for understanding
the blood flow behaviour, 
travelling wave solutions are still applicable by considering 
suitable boundary conditions.
Specifically, while travelling waves do not describe common morphologies
such as a fixed diameter or pressure, or a fixed blood flow, in a blood vessel,
nevertheless they may be relevant if
conditions in the vessel wall or surrounding tissue
cause a persistent wave pulse to propagate axially with constant speed.
They may also be relevant in constructing piecewise solutions
for approximating more realistic wave forms \cite{AnlMorOgd}.
Travelling waves also include steady-state (time-independent) solutions
as a special case when the wave speed is zero,
and these solutions are compatible with all standard morphological boundary conditions.

A variety of interesting behaviours exhibited by the travelling waves 
have been found, including:\\
$\bullet$ 
pressure shocks;\\
$\bullet$
blood flow shocks; \\
$\bullet$
sharp wave-front pulses in pressure and blood flow;\\
$\bullet$
flows in which the blood vessel is expanding or constricting.

All of the new results show the utility of symmetry analysis for providing 
explicit analytical information about exact non-steady solutions and conservation laws. 

There are several possible directions for future work: 
(1) understand piecewise solutions in a framework of weak solutions;
(2) examine stability of the solutions;
(3) derive and apply energy inequalities in the study of the initial-value problem;
(4) study similarity solutions using scaling and dilation symmetries;
(5) consider improved models, for example, by 
inclusion of a diffusion term, 
use of a viscoelastic tube law, 
and an improved radial velocity profile.

\section*{Acknowledgments}
SCA is supported by an NSERC Discovery Grant. 
APM and MLG warmly thank the research group FQM-201 
from the Andalusian Government for financial support. 
TMG acknowledges the \textit{Plan Propio - UCA} 2022-2023.



\end{document}